\documentclass[a4paper,12pt]{article}

\usepackage[utf8]{inputenc}
\usepackage[article]{mymath}
\usepackage[nottoc]{tocbibind}
\usepackage{marginnote}
\usepackage{enumitem}

\numberwithin{equation}{subsection}
\reversemarginpar

\title{Spinor analysis\\
{\normalsize (Spinoranalyse)\\ Nachrichten von der Gesellschaft der Wissenschaften zu Göttingen, 100\,--110,~(1929)}}

\author{ B.L. van der Waerden in Groningen (Holland)\\
\normalsize Submitted by R. Courant at the meeting on 26 July 1929\\
{\small(translated by Guglielmo Pasa\footnote{guglielmo.pasa@gmail.com}\ )}}

\date{{\small(\today)}}

\newcommand{\dr}{\dot{\rho}}
\newcommand{\dl}{\dot{\lambda}}
\newcommand{\dn}{\dot{\nu}}
\newcommand{\dm}{\dot{\mu}}

\newcommand{\Note}[1]{\footnote{{\color{blue}TN: {#1}}}}
\newcommand{\TN}[1]{\Note{  originally {#1}}}
\newcommand{\cfTN}[1]{\Note{ cf. note {#1}}}

\begin{document}
\begin{center}
\vskip3pc
{\Large Spinor Analysis}
\vskip1pc
{\large B.\@L.\@ van der Waerden in Groningen (Holland)}
\vskip1pc
Submitted by R.\@ Courant at the meeting on 26 July 1929
\vskip1pc
translated from
\vskip.5pc
{\small  Nachrichten von der Gesellschaft der Wissenschaften zu Göttingen, 100\,--110,~(1929)}
\vskip.5pc
by
\vskip.5pc
Guglielmo Pasa\footnote{guglielmo.pasa@gmail.com}
\end{center}

\tableofcontents
\vskip3cm
\marginnote{p.100}
"Let us call the novel quantities which, in addition to the vectors and tensors, have appeared in the quantum mechanics of the 
spinning electron, and which in the case of the Lorentz group are quite differently transformed from tensors, as spinors for 
short.
Is there no spinor analysis that every physicist can learn, such as tensor analysis, and with the aid of which all 
the possible spinors can be formed, and secondly, all the invariant equations in which spinors occur?" So Mr Ehrenfest 
asked me and the answer will be given below.

It turns out that the question can be solved without any new tools. It is necessary, in order to find all the possible 
spinors, to set up all the representations of the Lorentz group, and these are known\footnote{See, for example, H.\@ Weyl,
Group Theory and Quantum Mechanics, Leipzig, 1928, and the literature cited therein.}. In order to arrive at an 
invariant theory for the spinors, only the known two-step isomorphism between the Lorentz group and the binary 
unimodular group has to be converted into a "transfer principle" for the covariants of both 
groups\footnote{See E.\@ A.\@ Weiss, A Spatial Analogue to the Hessian Transfer Principle, Diss.\@ Bonn, 1924.},
and a known theorem of invariant theory In order to  see that all possible invariant equations can be written as binary tensor equations.

On the physical side, a complete overview of all possible invariant wave equations is obtained. It shows that a Lorentz 
invariant linear wave equation for the spinning electron with only two wave functions and of 
\marginnote{p.101}
the first order necessarily leads to the Dirac equation without the mass element, that is, is not consistent 
with the fact of the mass. It is therefore necessary to use either differential equations of the second order 
or at least four wave functions. This can be set up so: there are still a lot of possibilities.

\section{The representations of the Lorentz group.}
The task of finding all the "quantities" which are linearly transformed with Lorentz transformations according 
to some rule, so that the corresponding transformations of the "quantities" are also composed when the second Lorentz 
transformations are composed, ie the product of two Lorentz transformations is nothing but the problem of the 
representation of the Lorentz group by linear transformations.

It is known that the Lorentz group has a ambiguous representation as a binary group (that is, as a group in two complex variables); we will write this down first.

We start from a transformation group in 2 complex variables, which we will subsequently relate to the Lorentz group.

The binary transformations of determinant 1\TN{the index of the coefficients $\alpha$ in the transformations were all in lower position.}: 
\begin{equation*}\label{eq1}
(1)\quad\begin{cases}
\xi'_1&=\alpha_{1}{}^1\xi_1+\alpha_{1}{}^2\xi_2\\
\xi'_2&=\alpha_{2}{}^1\xi_1+\alpha_{2}{}^2\xi_2
\end{cases}
\qquad
(\bar{1})\quad\begin{cases}
\bar{\xi}'_1&=\bar{\alpha}_{1}{}^1\bar{\xi}_1+\bar{\alpha}_{1}{}^2\bar{\xi}_2\\
\bar{\xi}'_2&=\bar{\alpha}_{2}{}^1\bar{\xi}_1+\bar{\alpha}_{2}{}^2\bar{\xi}_2
\end{cases}
\end{equation*}
leave, when $\eta_1,\eta_2$ is transformed cogrediently to the $\xi_i$, the expression $\xi_1\eta_2-\xi_2\eta_1$ invariant, and therefore transform $\eta_2,-\eta_1$ 
contragrediently to $\xi_1,\xi_2$. We can therefore put 
\begin{equation}
\begin{cases}
\eta^1&=\eta_2\\
\eta^2&=-\eta_1
\end{cases}\tag{2}
\end{equation}
and also for all binary vectors and tensors, eg\TN{ the last expression reads: $\dots=a_{12}$.}
$$
\begin{cases}
\xi^1&=\xi_2\\
\xi^2&=-\xi_1
\end{cases}
\quad
\begin{array}{l}
a^{11}=a^1_{\ 2}=a_{22}\\
a^{21}=a^2_{\ 2}=-a_{12}.
\end{array}
$$
We agree to write a point (dotted indices) above the index in all variables which are transformed according to the 
conjugate complex transformation ($\bar{1}$)\TN{ ($\bar{4}$).}. The corresponding agreement applies to tensors; For example, $a_{\dl\dm\nu}$
is a tensor that transforms like $\bar{\xi}_\lambda\bar{\eta}_\mu\zeta_\nu$. For the dotted indices the above terms: 
$\xi^{\dot{1}}=\xi_{\dot{2}};\ \xi^{\dot{2}}=-\xi_{\dot{1}}$ are also valid.
\marginnote{p.102}

The expressions 
$$
\bar{\xi}_1\xi_1,\ \bar{\xi}_2\xi_2,\ \frac{\bar{\xi}_1\xi_2+\bar{\xi}_2\xi_1}{2},
\ \frac{\bar{\xi}_1\xi_2-\bar{\xi}_2\xi_1}{2i} 
$$
are real, and are transformed by the transformations ($1$), ($\bar{1}$)\TN{ ($4$), ($\bar{4}$).} into real ones again; 
So the transformation coefficients for these expressions are real. 
However, these same transformation coefficients also valid for a tensor $a_{\dot{a}\beta}$ and the expressions 
$$
a_{\dot{1}1},\ a_{\dot{2}2},\ \frac{a_{\dot{1}2}+a_{\dot{2}1}}{2},\ \frac{a_{\dot{1}2}-a_{\dot{2}1}}{2i}.
$$
Hence if 
$$
(3)\quad
\begin{cases}
\dfrac{a_{\dot{2}1}+a_{\dot{1}2}}{2}&=x\\ \\
\dfrac{a_{\dot{2}1}-a_{\dot{1}2}}{2i}&=y\\ \\
\dfrac{a_{\dot{1}1}-a_{\dot{2}2}}{2}&=z\\ \\
\dfrac{a_{\dot{1}1}+a_{\dot{2}2}}{2c}&=t
\end{cases}
\quad\text{or}\quad(4)\quad
\begin{cases}
a_{\dot{2}1}&=x+iy\\
a_{\dot{1}2}&=x-iy\\
a_{\dot{1}1}&=z+ct\\
-a_{\dot{2}2}&=z-ct
\end{cases}
$$ 
are transformed, then $x, y, z, t$ are also transformed as real. Moreover\footnote{Indices occurring above and below are 
summed up silently.}\TN{ the last line reads: $=a^{\dot{1}2}a_{\dot{2}1}-a^{\dot{2}2}a_{\dot{1}1}=x^2+y^2+z^2-c^2t^2.$}, 
\begin{equation}
\begin{aligned}
-\frac12a^{\dot{\alpha}\beta}a_{\dot{\alpha}\beta}
&=-\frac12(a^{\dot{1}2}a_{\dot{1}2}+a^{\dot{2}1}a_{\dot{2}1}+a^{\dot{1}1}a_{\dot{1}1}+a^{\dot{2}2}a_{\dot{2}2})\\
&=a_{\dot{1}2}a_{\dot{2}1}-a_{\dot{2}2}a_{\dot{1}1}=x^2+y^2+z^2-c^2t^2
\end{aligned}\tag{5}
\end{equation}
is invariant, so we have a real Lorentz transformation. It is also known that any real Lorentz transformation 
can be obtained in such a way\footnote{In order to see this, for example, it suffices to consider that a 6-membered set of 
Lorentz transformations is obtained from (1){\color{blue}(TN: originally (4))}.}. If we reverse the signs of $a_{\mu\nu}$ 
in (1)\cfTN{ $^5$}, we obtain 
the same transformation for the $a_{\dm\nu}$, that is, the transformations (1)\cfTN{ $^5$}.
form an double representation of the Lorentz group.
Finally, every transformation (1)\cfTN{ $^5$}  is to be obtained steadily from the identity, 
so, under our world transformations, neither the spatial reflections, nor the transformations, which change 
the course of the time, are preserved.

In order to generate the reflections, we also add to (1) the transformation with which every $\xi_\nu$ 
is transformed into  $\xi_{\dn}$\TN{ $\bar{\xi}_{\dn}$.}. 
It takes a form $a^{\dm\nu}\bar{\zeta}_{\dm}\xi_\nu$ in $a^{\mu\dn}\zeta_\mu\bar{\xi}_{\dn}$ and therefore $a^{\dm\nu}$ 
in $a^{\dn\mu}$ and also $a_{\dm\nu}$ in $a_{\dn\mu}$ over.
\marginnote{p.103}
It follows that $x, z$, and $t$ are invariant, but $y$ is transformed into $-y$, that is, it creates a reflection on the 
$XZ$ plane.

If we leave the reflections for the time being, it is clear that all representations of the group (1)\cfTN{ $^5$} are also 
representations of the Lorentz group, and vice versa.

However, the group (4) does not allow any other irreducible representations, as is shown in the representation theory, 
to be obtained when binary tensors $a_{\alpha\beta\dots\dot{\gamma}\dot{\delta}\dots}$, symmetric in $\alpha\beta\dots$ 
and $\dot{\gamma}\dot{\delta}\dots$, are used as transformation objects.
Any representation is composed of such irreducible, that is, every representation is obtained by taking one or more 
tensors $a_{\alpha\beta\dots\dot{\gamma}\dot{\delta}\dots}$ with or without symmetry conditions as the transformation 
object. 
Thus the "quantities" of the Lorentz group are no other than these binary tensors, transformed by the binary 
tensor representations of the Lorentz group. We call them 'the quantities of the Lorentz group', spin-tensors, or 
spinors, namely spin-tensors of the $1^\text{rst}$, $2^\text{nd}, \dots$ rank depending on the number of indices.
Specifically, the quantities with only one index  (assuming the values $1, 2$)  should be called spinvectors\footnote{ %
If one is not looking for the quantities of the Lorentz group but the three-dimensional group of rotations, 
it is easiest to consider this group as the subgroup of the Lorentz group, which leaves the time $t$ 
invariant.
The quantity $t$ transforms as $a_{\dot{1}1}+a_{\dot{2}2}$, thus also as a Hermite form 
$\bar{\xi}_1\xi_1+\bar{\xi}_2\xi_2$; the rotation group in the representation (1) is represented 
by the "unitary" subgroup, which leaves this form invariant.
In this subgroup, obviously, $\bar{\xi}_1,\bar{\xi}_2$ are transformed contragrediently to $\xi_1,\xi_2$, ie $\xi^1,\xi^2$.
Therefore the dotted indices become superfluous: they are replaced by contragredient undotted ones. 
The possible "quantities" which belong to the irreducible representations are once again the symmetric 
tensors $a_{\alpha\beta\dots\delta}$ of rank $2j = 0, 1, 2,\dots$}.

\section{Classification of world vectors and tensors.}
Since the binary tensor representations provide all the representations, the ordinary quaternary 
world vectors and tensors must also be found under it, that is, the world vectors must be written 
as spinors with binary indices.

A world vector can be given as $a^k$ (cogredient to $x, y, z, t$), or contragredient to it as $a_l=g_{lk}a^k$ 
($g_{00}=-c^2$, $g_{11}=1$ etc.). We introduce a new term for each world vector by setting (corresponding to (4)): 
\marginnote{p.104}
$$
\begin{cases}
-a^{\dot{1}2}&=a_{\dot{2}1}=a^1+ia^2=a_1+ia_2\\
-a^{\dot{2}1}&=a_{\dot{1}2}=a^1-ia^2=a_1-ia_2\\
a^{\dot{2}2}&=a_{\dot{1}1}=a^3+ca^0=a_3-\frac1ca_0\\
-a^{\dot{1}1}&=-a_{\dot{2}2}=a^3-ca^0=a_3+\frac1ca_0.
\end{cases}
$$

Accordingly, we designate all world tensors as binary tensors; for example, for a world tensor $a_{ik}$
(which is to be treated initially as a product $a_i b_k$)\TN{ the last line reads: $=a_{3,3}+\frac1c a_{3,0}+\frac1ca_{0,3}+\frac1{c^3}a_{0,0}$}:
\begin{alignat*}{1}
a_{\dot{2}1,\dot{2}1}&=a_{1,1}+ia_{2,1}+ia_{1.2}-a_{2,2}\\
-a_{\dot{2}1,\dot{1}2}&=a_{1,1}+ia_{2,1}-ia_{1,2}+a_{2,2}\\
\dots\\
a_{\dot{2}2,\dot{2}2}&=a_{3,3}+\frac1c a_{3,0}+\frac1ca_{0,3}+\frac1{c^2}a_{0,0}.
\end{alignat*}

The differentiation symbols 
$\frac{\partial}{\partial x} ,\frac{\partial}{\partial y},\frac{\partial}{\partial z},\frac{\partial}{\partial t}$ 
can also be renamed as they transform as $a_k$:
\begin{alignat*}{1}
\frac{\partial}{\partial x}+i\frac{\partial}{\partial y}&=\partial_{\dot{2}1}\\
\frac{\partial}{\partial x}-i\frac{\partial}{\partial y}&=\partial_{\dot{1}2}\\
\frac{\partial}{\partial z}-\frac1c\frac{\partial}{\partial t}&=\partial_{\dot{1}1}\\
\frac{\partial}{\partial z}+\frac1c\frac{\partial}{\partial t}&=-\partial_{\dot{2}2}.
\end{alignat*}

As can be seen from (5), 
$$
a_k a^k = g_{kl}a^ka^l=-\frac12a^{\dl\mu}a_{\dl\mu}
$$
and therefore also 
$$
a_kb^k=a^kb_k=g_{kl}a^kb^l=-\frac12a^{\dl\mu}b_{\dl\mu}=-\frac12a_{\dl\mu}b^{\dl\mu}.
$$
Of course, these formulas also apply if the vectors $a, b$ are replaced by differentiation symbols;
for example, 
\begin{alignat*}{1}
\mathrm{div}\ a &= -\frac12\partial_{\dm\nu}a^{\dm\nu}\\
\square
&=\sum\frac{\partial^2}{\partial x_2^2}-\frac1{c^2}\frac{\partial^2}{\partial t^2}
=-\frac12\partial_{\dl\mu}\partial^{\dl\mu}.
\end{alignat*}

The following table lists the possible types of spinors of the lowest grades, indicating which species of 
world tensors correspond to them:
\marginnote{p.105}

\begin{center}
\begin{tabular}{l | l}
& Spinors\\
\hline
rank $1$ & $a_{\dl},\ a_\lambda$ (spin-vectors),\\[.5pc]
rank $2$ & $a_{\lambda\mu},\ a_{\dl\dm},\ a_{\dl\mu}\rightleftarrows$ world vector $a_k$,\\[.5pc]
rank $3$ & $a_{\lambda\mu\nu},\ a_{\dl\dm\dn},\ a_{\lambda\mu\dn},\ a_{\dl\dm\nu}$,\\[.5pc]
rank $4$ & $\begin{cases}
a_{\lambda\mu\nu\rho},\ &a_{\dl\dm\dn\dr},\ a_{\lambda\mu\nu\dr},\ a_{\dl\dm\dn\rho}\\
a_{\dl\mu\dn\rho} \rightleftarrows & \text{world tensor }a_{kl}.
\end{cases}$
\end{tabular}
\end{center}

It should also be noted that one can also allow the spinors of the second rank $a_{\lambda\nu}$ and $a_{\dl\dm}$ 
to correspond to world tensors, for example the "selfdual" $F_{kl}$, by setting: 
$$
a^{\alpha}_{\beta\dl\dm}=\delta^\alpha_\beta a_{\dl\dm}
$$
from which 
$$
a_{\dl\dm}=\frac12a^\alpha_{\alpha\dl\dm}
$$
follows and by forming the corresponding $a_{\alpha\beta\dl\dm}$ from $F_{kl}$.

It should also be pointed out that one can always restrict oneself to such spin tensors, which are 
symmetric in all the dotted and also in all the undotted indices, since all the others are formed linearly from them
with the aid of $\epsilon$- or $\delta$-symbols (see below).

\section{Invariant theory of the spinors.}
We consider a system of binary vectors and tensors, tentatively without dotted indices. 
All invariants and covariants of this system in the group (1)\cfTN{ $^5$} are obtained by a known 
theorem of binary  invariant theory ("the first fundamental principle of the symbolic method"): 
all the indices of the tensors are  written down and form expressions like 
$$
\epsilon^{\alpha\beta}\epsilon^{\gamma\delta}a_{\alpha\beta\gamma}b_{\delta\lambda\mu}\dots,
\quad\text{where}\ 
\begin{cases}
\epsilon^{12}=1,\ &\epsilon^{21}=-1,\\
\epsilon^{11}=\epsilon^{22}&=0
\end{cases}.
$$
The calculation rules for the symbol $\epsilon$ are
\begin{equation}
\begin{cases}
\epsilon^{\alpha\beta}&=-\epsilon^{\beta\alpha}\\
\epsilon^{\alpha\beta}\epsilon^{\gamma\delta}+\epsilon^{\beta\gamma}\epsilon^{\alpha\delta}+\epsilon^{\gamma\alpha}\epsilon^{\beta\delta}&=0\\
\epsilon^{\alpha\beta}u_{\alpha\beta\gamma}+\epsilon^{\alpha\beta}u_{\gamma\alpha\beta}+\epsilon^{\alpha\beta}u_{\beta\gamma\alpha}&=0.
\end{cases}\tag{6}
\end{equation}
Another slightly shorter type of writing for the invariants is obtained by using the rule (2) for raising the 
\marginnote{p.106}
indices and eliminating the $\epsilon$-symbols by means of: 
$$
\epsilon^{\alpha\beta}a_{\alpha\beta}=a_\alpha{}^\alpha=-a^\alpha{}_\alpha\footnote{The reason why the even shorter and more convenient symbolic factorization of the tensors has been avoided here consists chiefly in the difficulty of the designation of the differentiation symbols, where it is difficult to determine the magnitudes to which they are to act.}.
$$

The calculation rules we need now are harder to remember than (6) (from which follow):
\begin{equation}
\begin{cases}
u_\alpha{}^\alpha &=- u^\alpha{}_\alpha\\
u^\beta{}_{\beta\alpha}&=u^\beta{}_{\alpha\beta}-u_\alpha{}^\beta{}_\beta.
\end{cases}\tag{7}
\end{equation}
($1$) and ($\bar{1}$) are algebraically completely independent, and the invariance of a system of equations 
between binary vectors is expressed by the requirement of invariance for two independent groups, one of which 
operates only on the dotted indices, and the other only on the undotted indices.
Thus exactly the same invariant operations are permissible as before, but they must refer either to the 
dotted or to the undotted indices. That is, a raised, dotted index remains dotted, and is summed only by two dotted or two undotted indices.

If a system of equations is to be invariant under reflections, it must allow the replacement of all $a^{\dm\nu}$ by 
$a^{\dn\mu}$, and all $\xi_{\mu}$ by $\xi_{\dm}$\TN{ $\bar{\xi}_{\dm}$.} (or quantities which likewise transform).

\section{The Dirac wave equation}
The Dirac wave equation of the electron is multiplied by the Dirac $\Gamma_0$:
\begin{equation}
\frac1c\left(\frac hi\frac{\partial}{\partial t}+\Phi_0\right)\psi+\sum_1^3s_r'\left(\frac hi\frac{\partial}{\partial x_r}+\Phi_r\right)\psi+mc\Gamma_0\psi=0\footnote{The difference from Weyl §39, p. 172 in the member with $\Phi_0$
is that we use $x, y, z, t$, and not $x, y, z, ict$ as coordinates; Our vector $\Phi_0$ is real and contragredient to the coordinates and our $\Phi_0$ is $= ci$ times the Weyl one.}\tag{8}
\end{equation}
 where the $s'_r$ and $\Gamma_0$ are four-row matrices, 
 $$
 s_r'=\begin{pmatrix}
 s_r & 0\\
 0 & -s_r
 \end{pmatrix},
 \qquad
 \Gamma_0=\begin{pmatrix}
 0 & E\\
 E & 0
 \end{pmatrix},
 $$
where $E$ is the two-row unit matrix and the $s_r$ are the Pauli 
\marginnote{p.107}
matrices
$$
s_1=\begin{pmatrix}
0 & 1\\ 1 & 0
\end{pmatrix},
\quad
s_2=\begin{pmatrix}
0 & -i\\ i & 0
\end{pmatrix},
\quad
s_3=\begin{pmatrix}
1 & 0\\ 0 & -1
\end{pmatrix}.
$$
We denote the first two components of the Dirac wave function $\psi$ with $\psi_{\dot{1}}$, $\psi_{\dot{2}}$; the 
last two (transforming as the conjugate complex and  contragrediently to the first) by $\chi^1$, $\chi^2$, and (8) 
accordingly split into two equations: 
$$
\begin{cases}
\frac1c\left(\frac hi\frac{\partial}{\partial t}+\Phi_0\right)\psi+\sum_1^3 s_r\left(\frac hi\frac{\partial}{\partial x_r}+\Phi_r\right)\psi+mc\chi&=0.\\
\frac1c\left(\frac hi\frac{\partial}{\partial t}+\Phi_0\right)\chi-\sum_1^3 s_r\left(\frac hi\frac{\partial}{\partial x_r}+\Phi_r\right)\chi+mc\psi&=0.
\end{cases}.
$$
If we introduce the abbreviation $a_k$ for $\frac hi\frac{\partial}{\partial x^k}+\Phi_k$\TN{ $\frac hi\frac{\partial x^k}{\partial}+\Phi_k$.}, then we have to evaluate 
the quantity
$$
\frac1c a_0+\sum_1^3s_ra_r
$$
in the first equation.
It is the matrix 
$$
\begin{pmatrix}
\frac1ca_0+a_3 & a_1-ia_2\\
a_1+ia_2 & \frac1c a_0-a_3
\end{pmatrix}
=
\begin{pmatrix}
-a^{\dot{1}1}& -a^{\dot{2}1}\\-a^{\dot{1}2}&-a^{\dot{2}2}
\end{pmatrix}
=-
\begin{pmatrix}
a^{\dot{1}1}& a^{\dot{2}1}\\a^{\dot{1}2}&a^{\dot{2}2}
\end{pmatrix}.
$$
Also, in the second equation, the expression 
$$
\frac1ca_0-\sum_1^3s_ra_r=
\begin{pmatrix}
\frac1ca_0-a_3 & -a_1+ia_2\\
-a_1-ia_2 & \frac1ca_0+a_3
\end{pmatrix}
=
\begin{pmatrix}
-a^{\dot{2}2}& -a^{\dot{2}1}\\-a^{\dot{1}2}&-a^{\dot{1}1}
\end{pmatrix}
=
-
\begin{pmatrix}
a_{\dot{1}1}& a_{\dot{1}2}\\a_{\dot{2}1}&a_{\dot{2}2}
\end{pmatrix}\TN{ $\dots=\begin{pmatrix}
-a^{\dot{1}1}& -a^{\dot{1}2}\\-a^{\dot{2}1}&-a^{\dot{2}2}
\end{pmatrix}
=
-
\begin{pmatrix}
a^{\dot{1}1}& a^{\dot{1}2}\\a^{\dot{2}1}&a^{\dot{2}2}\end{pmatrix}$}
$$
is applied. Hence the Dirac equations are written as:
$$
\begin{cases}
-a^{\dl\mu}\psi_{\dl}+mc\chi^\mu&=0\\
-a_{\dm\lambda}\chi^\lambda+mc\psi_{\dm}&=0
\end{cases}
$$
or, if we lower the indices of the $\chi$ and use the meaning for the $a_{\dl\mu}$: 
\begin{equation}
\begin{cases}
-\left(\frac hi\partial^{\dl}{}_\mu+\Phi^{\dl}{}_\mu\right)\psi_{\dl}+mc\chi_\mu&=0\\
\left(\frac hi\partial_{\dm}{}^{\lambda}+\Phi_{\dm}{}^{\lambda}\right)\chi_\lambda+mc\psi_{\dm}&=0
\end{cases}\tag{9}
\end{equation}

\marginnote{p.108}
The derivation of the second-order wave equation 
\begin{equation}
h^2\square\chi_\nu=-\frac12h^2\partial^{\dl\mu}\partial_{\dl\mu}\chi_\nu=m^2c^2\chi_\nu\quad(\text{in the case }\Phi_k=0)\tag{10}
\end{equation}
(and also for $\psi$) from the Dirac occurs by using the identity (7): 
\begin{alignat*}{1}
\partial^{\dl\mu}\partial_{\dl\mu}\chi_\nu
&=\partial^{\dl\mu}\partial_{\dl\nu}\chi_\mu-\partial^{\dl}{}_\nu\partial_{\dl}{}^\mu\chi_\mu
=-2\partial^{\dl}{}_\nu\partial_{\dl}{}^\mu\chi_\mu\\
&=\frac{2mci}{h}\partial^{\dl}{}_\nu\psi_{\dl}=\frac{2m^2c^2i^2}{h^2}\chi_{\lambda}.
\end{alignat*}

\section{The possible Lorentz invariant wave equations.}
From a wave equation for the electron we have to demand that it is linear in the occurring wave functions 
$\psi$ and that it expresses the derivative $\frac{\partial\psi}{\partial t}$ or $\frac{\partial^2\psi}{\partial t^2}$ 
linearly on the other derivatives of the same and lower order, with coefficients which may still be appended by the 
field.
If the wave function $\psi$ has two components $\psi_1$, $\psi_2$, then the wave equation must also consist of two 
components.

If one also requires that it is of the first order, then only 
$$
\partial_{\dl\mu}\psi^\mu+c_{\dl\mu}\psi^\mu=0
$$
can be considered where the $c_{\dl\mu}$ may depend on the field. For ``no field'' we must have $c_{\dl\mu}=0$, 
and one obtains a pair of the Dirac equations with $m=0$.

If we let four $\psi$-components $m$, of which the first two are transformed according to (1), the last two to ($\bar{1}$), 
one obtains as possibilities: 
$$
\begin{cases}
\partial_{\dl\mu}\psi^\mu+b_{\dl\mu}\psi^\mu+c_{\dm}^{\dot{\sigma}}\psi_{\dot{\sigma}}&=0\\
\partial_{\dl\mu}\psi^{\dl}+e_{\dl\mu}\psi^{\dl}+f^\rho_\mu\psi_\rho&=0.
\end{cases}
$$
For ``no field'', only one multiple of the unit matrix remains for $c^{\dl}_{\dm}$ and $f^\rho_\mu$, and the Dirac 
equations are obtained by giving the opposite sign to these multiples. For the field dependence of $b, c, e, f$ there are, 
of course, more possibilities than the Dirac alone.
\marginnote{p.109}
If a differential equation of the second order is obtained, with only two components of the wave function, the number of possibilities is still much greater: the following terms in the wave equation are possible: 
$$
\square\psi^\mu+b^{\dl\rho\mu}_\sigma\partial_{\dl\rho}\psi^\sigma+c_\lambda^\mu\psi^\lambda=0.
$$
(The conceivable $\partial^{\dr\mu}\partial_{\dr\nu}\psi^\nu$ can be expressed by $\square\psi^\mu$: cf. § 4, conclusion.)
\vskip1cm
\centerline{\rule{5cm}{1pt}}
\vskip2cm
\appendix
The following appendix is not present in the actual paper, but is provided here as a terminology complement.
\section{Contragredient, cogredient transformations}\label{appA}
Let $A$ be the transformation matrix
$$
A=
\begin{pmatrix}
\alpha_{1}{}^1 & \alpha_{1}{}^2\\ \alpha_{2}{}^1 & \alpha_{2}{}^2
\end{pmatrix}
$$
The vectors $X$ and $Y$ transform {\em cogrediently} if
$$
X'=AX,\ \text{and}\ Y'=AY\quad \Leftrightarrow\ x_\mu'=\alpha_\mu{}^\nu x_\nu\ \text{and}\ y_\mu'=\alpha_\mu{}^\nu y_\nu
$$
in which case
$$
{X'}^TY'=X^TA^TAY.
$$

The vectors $X$ and $Y$ transform {\em contragrediently} if
$$
X=A^TX'\Leftarrow\ X'=(A^{-1})^TX,\ \text{and}\ Y'=AY\quad\Leftrightarrow\quad x_\mu=x_\nu'\alpha^\nu{}_\mu\ 
\text{and}\ y_\mu'=\alpha_\mu{}^\nu
$$
($\alpha^\nu{}_\mu$ is the transpose matrix of $\alpha_\mu{}^\nu$) in which case
$$
X'{}^TY'=XA^{-1}AY=X^TY.
$$
\section{Remarks}
This translation has been made available, as is, for whoever is interested, even though it was made for the personal 
use of the author. If you find any mistake or better expressions, feel free to contact the author of the translation 
for an update.

The footnotes numbering in the original publication were made on a per page basis. Here, instead, it is continuous and 
the original footnotes are interspersed with translation comments (see below), so the original footnotes numbering is 
not preserved.

I've tried to correct the typos in the original paper. Corrections and translation remarks are preceded 
by `{\color{blue}TN:}' in footnotes and those footnotes are displayed in {\color{blue}blue} color. All other footnotes 
come from the original paper.

The transformation matrix elements of equations (1) and ($\bar{1}$) are displayed with one upper index and one lower index 
to make notations coherent with modern implicit summation convention (one upper index and one lower equal index letter 
are meant to be summed on).

The appendix \ref{appA} has been included to make clear the meaning of these expressions.
\end{document}